\documentclass[usenatbib]{mnras}

\usepackage{graphicx}
\usepackage{gensymb}
\usepackage{booktabs}
\usepackage[T1]{fontenc}        
\usepackage[utf8]{inputenc}
\usepackage{babel}
\usepackage{subfigure}
\usepackage[font=small,labelfont=bf,tableposition=top]{caption}
\usepackage{threeparttable}  
\usepackage{makecell}
\usepackage{multirow}
\usepackage{diagbox}
\usepackage{stackengine}
\usepackage{amssymb}

\usepackage{newtxmath}

\usepackage[T1]{fontenc}

\DeclareRobustCommand{\VAN}[3]{#2}
\let\VANthebibliography\thebibliography
\def\thebibliography{\DeclareRobustCommand{\VAN}[3]{##3}\VANthebibliography}

\usepackage{graphicx}	
\usepackage{amssymb}	

\title[RFI Identification Based On Deep-Learning]{A Robust RFI Identification For Radio Interferometry based on a Convolutional Neural Network}

\author[Haomin Sun et al.]{
 Haomin Sun$^{1,2}$,
 Hui Deng$^{1,2}$\thanks{E-mail: denghui@gzhu.edu.cn},
 Feng Wang$^{1,2}$\thanks{E-mail: fengwang@gzhu.edu.cn},
 Ying Mei$^{1,2}$,
 Tingting Xu$^{1,2}$, 
 \newauthor
 Oleg Smirnov$^{3}$,
 Linhua Deng$^{4}$, 
 and Shoulin Wei$^{5}$
\\
$^{1}$ Center For Astrophysics, Guangzhou University, Guangzhou 510006, P.R. China\\
$^{2}$ Great Bay Center, National Astronomical Data Center, Guangzhou, Guangdong, 510006, P.R. China\\
$^{3}$ Department of Physics and Electronics, Rhodes University, PO Box 94, Makhanda 6140, South Africa\\
$^{4}$ Yunnan Observatory, Chinese Academy of Sciences, Kunming, Yunnan, 650216, P.R. China\\
$^{5}$ Key Lab Of Computer Technology Appliance, Kunming University of Science And Technology, Kunming, Yunnan, 650500, P.R. China\\
}

\date{Accepted XXX. Received YYY; in original form ZZZ}

\pubyear{2021}

\begin{document}
\label{firstpage}
\pagerange{\pageref{firstpage}--\pageref{lastpage}}
\maketitle

\begin{abstract}
The rapid development of new generation radio interferometers such as the Square Kilometer Array (SKA) has opened up unprecedented opportunities for astronomical research. However, anthropogenic Radio Frequency Interference (RFI) from communication technologies and other human activities severely affects the fidelity of observational data. It also significantly reduces the sensitivity of the telescopes. We proposed a robust Convolutional Neural Network (CNN) model to identify RFI based on machine learning methods. We overlaid RFI on the simulation data of SKA1-LOW to construct three visibility function datasets. One dataset was used for modeling, and the other two were used for validating the model's usability. The experimental results show that the Area Under the Curve (AUC) reaches 0.93, with satisfactory accuracy and precision. We then further investigated the effectiveness of the model by identifying the RFI in the actual observational data from LOFAR and MeerKAT. The results show that the model performs well. The overall effectiveness is comparable to AOFlagger software and provides an improvement over existing methods in some instances.

\end{abstract}



\section{Introduction}
\label{1}

During the past two decades, multiple radio telescopes have been commissioned or scheduled to be built.
Several pathfinders of the Square Kilometer Array (SKA) \citep{Braun+1996}, i.e., the Murchison Widefield Array (MWA)~\citep{lonsdale2009murchison}, the Australian SKA Pathfinder (ASKAP)~\citep{macquart2010commensal}, the South African MeerKAT~\citep{booth2012overview}, and the Low Frequency Array (LOFAR)~\citep{de2009lofar}, have been put into routine observation, laying the foundation for the construction of the SKA.

Any unwanted signal in radio astronomy is considered Radio Frequency Interference (RFI), regardless of whether it is intentional or not. RFI affects all radio astronomy telescopes and often originates in intentional emitters such as digital TV, mobile and satellite communications systems etc.
\citep{offringa2015low} showed that DTV, radio, and satellite communication had become significant sources of local RFI.  In general, flagging the visibility function data affected by RFI is necessary for modern radio interferometers. Incorrectly flagged visibility data causes incorrect imaging results~\citep{perley2003evla}. 

Identification and mitigation of RFI has attracted a substantial amount of research work. The basic principle is to set a threshold for the amplitude of the data to detect RFI~\citep{parsons2014new}. AOflagger, a tool for mitigating RFI used by LOFAR, adopted the SumThreshold method~\citep{offringa2010lofar}. RFIMS method was proposed for the Westerbork Synthesis Radio Telescope is based on a powerful field-programmable gate array processor~\citep{baan2004radio}. The polarization information of post-correlation interference signals is used to detect and mitigate RFI~\citep{yatawatta2021polarization}. These methods have significantly improved the development of high-precision radio observations with outstanding results. However, SKA science goals have led to new challenges for traditional RFI identification and mitigation methods. For example, exploring Epoch of Reionization (EoR) is one of the most important scientific goals of the SKA, which requires a data reduction system to identify and mitigate faint radio frequency interference (i.e., low-level RFI)~\citep{offringa2019impact}. 

It is worth exploring some new approaches to solve the RFI problem faced by radio interference arrays. 
Deep learning is being used widely and extensively for classification problems in recent years. Convolutional neural networks (CNN) are being adopted in image recognition. U-Net enables the classification of clean signal and RFI signatures in 2D time-ordered data by using a particular type of CNN \citep{akeret2017radio}. R-Net algorithm and transfer learning seem to be a practical algorithm for mitigating RFI~\citep{vafaei2020deep}. The deep fully convolutional neural network works well by using both amplitude and phase information to identify RFI~\citep{kerrigan2019optimizing}. The algorithm for identifying RFI based on deep learning (e.g., U-Net) often requires a large amount of RFI data to train the neural network, and its accuracy also depends on whether a correct label map is given to the neural network~\citep{vafaei2020deep}.

Our research focused on the application of CNN on RFI identification. We studied RFI simulation methods, constructed a CNN model for the visibility functions simulated from the SKA1-LOW configuration, and applied the model to the real observational data from two radio telescopes. The rest of this study is organized as follows.

Sec. \ref{2} constructs a CNN model for the RFI identification. Sec. \ref{3} describes the simulation process, including visibility data simulation and RFI simulation, checks out CNN model on the simulated data based on the configuration of SKA1-LOW. Sec. \ref{4}, applies our model to the real observational data from MeerkAT and LOFAR. Sec. \ref{5} evaluates our CNN model and discusses some issues. Sec. \ref{6} gives conclusions.

\section{Method}
\label{2}

\subsection{Data}
\label{method}
The radio interferometer reconstructs spatial and spectral information from incomplete visibilities sampled at multiple observational frequencies. An interferometric array provides samples of the complex visibility functions of the sources in the $(u,v)$ plane\citep{cornwell1999deconvolution}. The visibility function $V(u,v)$ is related to the source intensity distribution $I(l,m)$ (multiplied by the primary beam of the array elements) by a two-dimensional Fourier transform:

\begin{equation}\label{fun1}
V(u, v)=\iint_{S} I(l, m) e^{-2 \pi i(u l+v m)} d l d m 
\end{equation}

where $l$ and $m$ are east-west, and north-south angles on the sky, $u$ and $v$ are east-west and north-south spatial frequencies.

\begin{figure}
\begin{center}
	\includegraphics[width=\columnwidth]{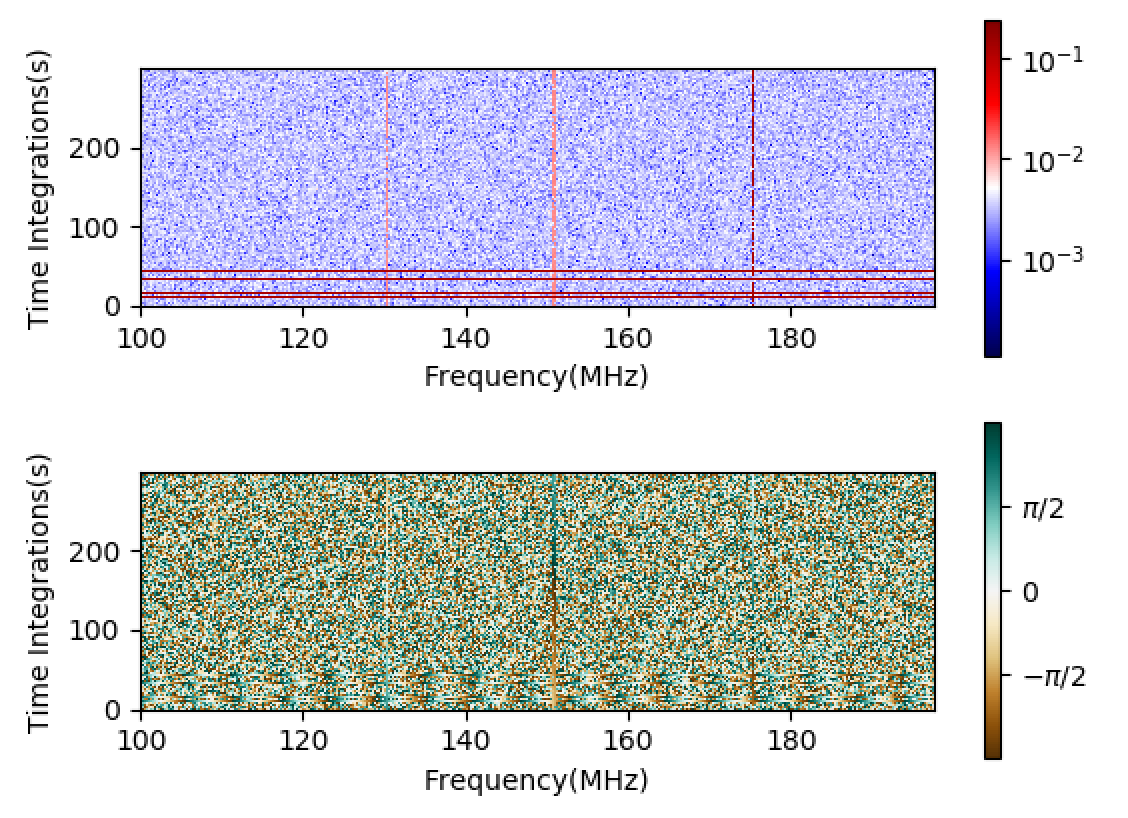}
\end{center}
    \caption{An example of SKA1-LOW visibility simulated from baseline SKA1-LOW020 - SKA1-LOW004. The top panel shows the amplitude. The bottom panel shows the phase. The two maps provide some information about the RFI, which has constant narrow-band RFI in the channel of 130MHz, 152MHz, and 178MHz, and broad-band RFI happens from 10 to 40 seconds.}
    \label{fig:RFI}
\end{figure}

From the visibility data, there is a difference between the astronomical signal affected by RFI and the normal astronomical signal
~\citep{baan2011rfi}. Fig.~\ref{fig:RFI} shows the amplitude and phase of the visibility data for time vs. frequency. It is evident find that the data without RFI contains stable values in amplitude and phase, while the RFI-Polluted data contains values that change significantly. 

When using CNN patterns to identify and mitigate RFI, we must first construct the underlying dataset so that the CNN can learn features that exist in the amplitude and phase information of the visibility data, then transform the identification of RFIs into a binary classification. We took the amplitude and phase values to form a data pair, and all the data pairs constituted the dataset for the following CNN modeling.

\subsection{CNN Design}


We proposed a dichotomous model to identify RFI after analyzing the data features with RFI. The model is inspired by a classic architecture named Le-Net architecture  \citep{lecun1998gradient}, which has a good performance in many domains such as image classification \citep{el2016cnn} , pattern recognition \citep{yuan2012offline} and computer vision \citep{xie2016disturblabel}. 

The model consists of 4 convolutional layers, 4 Max-pooling layers, and 2 fully connected layers.
For the first two convolutional layers, we chose a convolutional kernel of $1\times2$ and used one as the parameter for striding and padding. The remaining convolutional layers have a convolutional kernel size of $3\times3$. The parameters of striding and padding are the same as that of the first two convolutional layers. For each convolutional layer, we chose Batch Normalization (BN) \citep{DBLP:journals/corr/IoffeS15} as regularization to speed up the training rate of the model. We chose Rectified Linear Unit (ReLU) \citep{10.5555/3104322.3104425} which is the most popular activation function for the activation layer which can generate a non-linear mapping between the input and output.
We added a max-pooling layer after each of the other convolutional layers, excluding the first convolutional layer. In the max-pooling layer, the kernel size of the first two layers are 1$\times$2, and the kernel sizes of the last two layers are 2$\times$2 and 3$\times$3, respectively.

\begin{figure*}
\begin{center}
	\includegraphics[width=17cm]{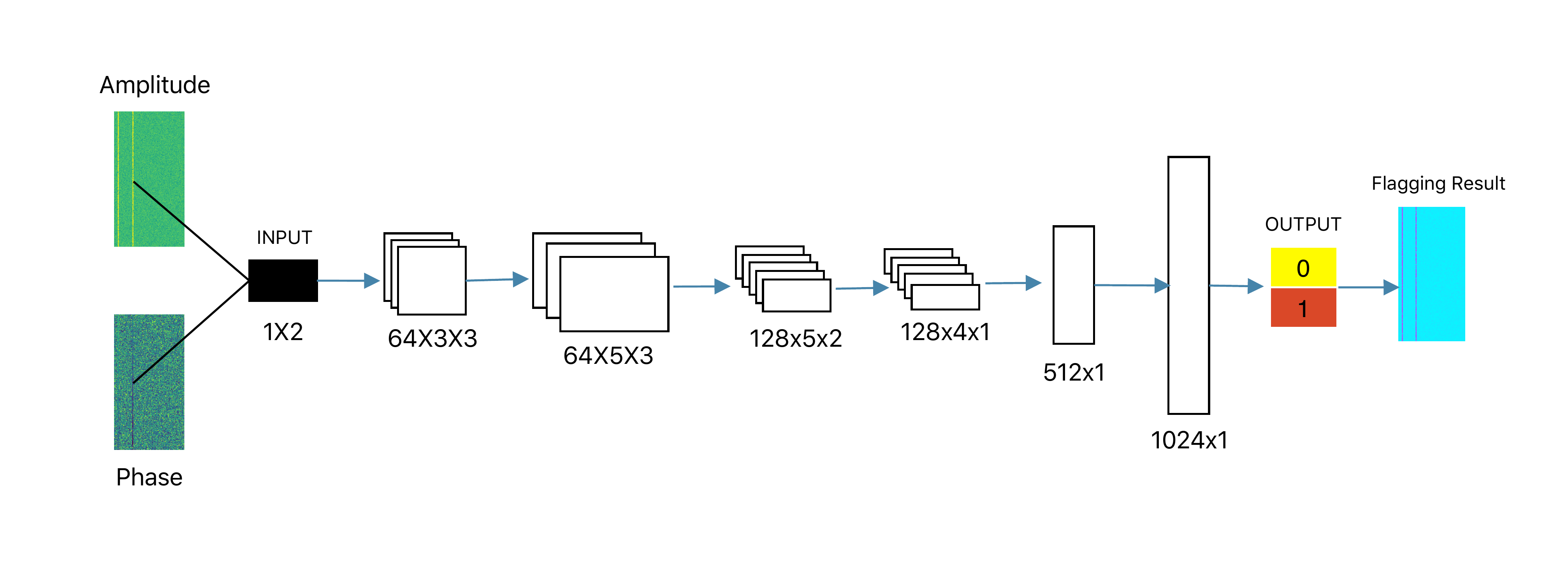}
\end{center}
    \caption{Schematic diagram of the CNN model architecture. }
    \label{fig:net}
\end{figure*}

After the features were obtained through convolutional processing, we added fully connected layers to map the learned "distributed feature representation" to the sample labeling space. For our first fully connected layer, we chose Dropout \citep{6179821} as regularization to prevent over-fitting and ReLU as the activation function. The final classification results were output by the classifier softmax in the last fully connected layer. The network is shown in Fig.~\ref{fig:net}, and the detailed information is listed in Table \ref{tab:Le-Net}.

\begin{table*}
\centering

\caption{Architecture overview of the CNN model proposed in the study.\label{tab:Le-Net}}
\begin{tabular}{ccccccccc}
\toprule
Layer & Type & Channels & Kernel Size & Striding & Padding & Regularization & Activation \\
\midrule
IN & INPUT & 1 & 1$\times$2 & 1 & 1 & BN & ReLU  \\
C2 & Convolution & 64 & 1$\times$2 & 1 & 1 & BN & ReLU \\
M3 & Max Pooling & 64 & 1$\times$2 & 1 & 0 & ... & ... \\
C4 & Convolution & 64 & 3$\times$3 & 1 & 1 & BN & ReLU  \\
M5 & Max Pooling & 64 & 1$\times$2 & 1 & 0 & ... & ... \\
C6 & Convolution & 128 & 3$\times$3 & 1 & 1 & BN & ReLU  \\
M7 & Max Pooling & 128 & 2$\times$2 & 1 & 0 & ... & ... \\
C8 & Convolution & 512 & 3$\times$3 & 1 & 1 & BN & ReLU  \\
M9 & Max Pooling & 512 & 3$\times$3 & 2 & 0 & ... & ... \\
FL10 & FlattenLayer & 512 & ... & ... & ... & Dropout & ReLU \\
OUT & FlattenLayer & 1024 & ... & ... & ... & ... & Softmax \\

\bottomrule
\label{modt}
\end{tabular}
\end{table*}

The model has been extensively tested and optimized and is the best of all model networks in the study. 
We initially used convolution kernels of $1\times$1 and $1\times$2 to build the network. However, it was eventually found that the feature map had to be extended using the padding method to effectively extract the input data ($1\times$2) features.

The number of layers and normalizations of the model are obtained after many trials. The trial results show that the detection efficiency of RFI with a four-layer network is optimal to maximize the learning time and minimize the prediction time and enable the network to learn as much as possible the abstract features of RFI.
The hyperparameters are also guided by the test results obtained from extensive trials.

\subsection{Model training and testing}

For a given dataset, we set the batch size as 16 and iteration time as 100, selected 16 samples each time to train the model. In the end, We selected the best performance model.

Each training sample was given a prediction label after training, and the loss was obtained by feeding the prediction label and true label into the loss function. We updated the parameters of the model by backpropagation to obtain the minimum loss. We used the gradient descent algorithm to update the parameters. The final loss value converged
to some constant values through iterative training. We chose the SGD (Stochastic Gradient Descent) \citep{2012Efficient} method for the momentum update and set the SGD parameters to the specified values, where the learning rate was 0.001 \citep{wilson2003general}, and the momentum was 0.9 \citep{sutskever2013importance}. Both of these values are standard values obtained from a large number of previous experiments in the field of deep learning.

After training, we used a test set to test the model and calculated evaluation metrics of the model.

\subsection{Model Evaluation}
Recall, Accuracy, and Precision are used to evaluate the results of the Le-Net. The three metrics are defined as follows:

\begin{equation}
\mathrm{Recall}=\frac{\mathrm{TP}}{\mathrm{TP}+\mathrm{FN}}
\end{equation}

\begin{equation}
\mathrm{Precision}=\frac{\mathrm{TP}}{\mathrm{TP}+\mathrm{FP}}
\end{equation}

\begin{equation}
\mathrm{Accuracy}=\frac{\mathrm{TP}+\mathrm{TN}}{\mathrm{TP}+\mathrm{FN}+\mathrm{FP}+\mathrm{TN}}
\end{equation}
where $TP$, $TN$, $FP$, $FN$ represent True Positive, True Negative, False Positive, and False Negative, respectively. 

It is important to note that the number of each type of dataset was unbalanced, which would result in a decrease of the performance of the Le-Net model~\citep{bhowan2012evolving}. Therefore, we plotted the ROC (Receiver Operating Characteristic) curve, computed the AUC (Area Under roc Curve)~\citep{bradley1997use} value and took AUC as a metric to evaluate our CNN model. 

\section{SKA1-LOW RFI Mitigation}
\label{3}

SKA1-LOW~\citep{mellema2013reionization} is the low-frequency~(50-350 MHz) telescope of the SKA Phase 1, to be constructed in Western Australia. There are 512 stations distributed mainly within a dense area around the center of the SKA1-LOW, and each station consists of 256 log-periodic dual-polarized antennas. It will be used to study the early epochs of the universe by looking at the red shifted signal from the 21 cm Hydrogen emission line. The reason for using the SKA1-LOW instead of the SKA1-MID for the research of the RFI mitigation is that the low-frequency interferometer is more susceptible to interference from RFI signals and presents more difficulties for RFI mitigation. 
This is due to the low-frequency bands being heavily in use by many communication systems. That results in strong signals all over the bands \citep{baan2011rfi}.

\subsection{SKA1-LOW Data Preparation}



\subsubsection{Visibility Data Simulation}
\label{norfi}
We developed a simulation software for the study based on the SKA Radio Astronomy Simulation, Calibration and Imaging Library (RASCIL)~\footnote{https://gitlab.com/ska-telescope/external/rascil} that expresses radio interferometry calibration and imaging algorithms in Python and NumPy. The SKA SIM/ORCA agile team has developed a series of software for SKA1-LOW and SKA1-MID, verifying RASCIL’s accuracy and robustness.

\begin{figure}
    \centering
	\includegraphics[width=6.5cm]{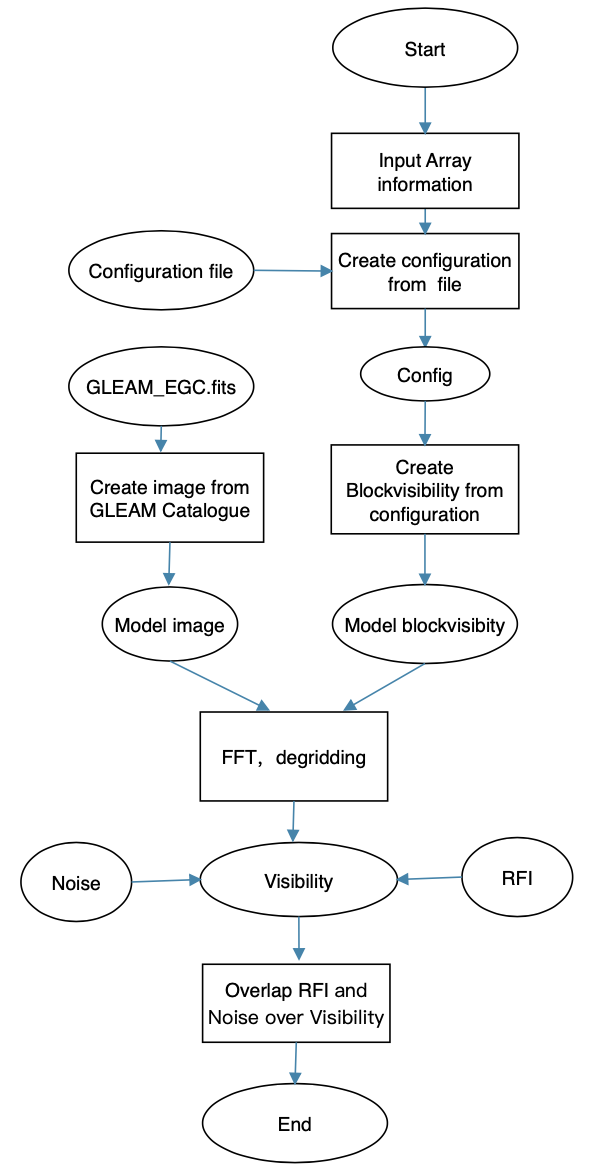}
    \caption{The flow chart for SKA1-LOW simulation observation}
    \label{fig:flow}
\end{figure}

We simulated observations of the SKA1-LOW telescope based on the telescope configurations given in the SKA design baseline document~\citep{dewdney2019ska1}. The flow chart of the simulation is shown in Fig.~\ref{fig:flow}. At the same time, we used the Extragalactic All-sky MWA~(GLEAM) catalog~\citep{wayth2015gleam}  to simulate the observation of point sources because the MWA telescope is one of the pathfinders for the SKA and its observational band and location are similar to that of SKA1-LOW. The Gleam survey contains 307,455 radio sources with 20 separate flux density measurements across 72 - 231 MHz~\citep{wayth2015gleam}. 

We generated three datasets with different periods. We simulated observations for 1000 channels ranging from 100 MHz to 200 MHz. The observation time is 10 minutes. The bandwidth of each channel is 0.1MHz. We only generated the visibility functions of Stokes I and stored the results in a Numpy array with a shape of (300,31,31,1000,1) that corresponds to (Time, Antenna1, Antenna2, Frequency, Polarization). 
Ultimately, we stored the three datasets in three Measurement Sets, i.e., Sim\_RFI-1, Sim\_RFI-2, and Sim\_RFI-3, respectively.

\subsubsection{RFI Simulation}
\label{rfi}

We adapted the hera\_sim\footnote{https://github.com/HERA-Team/hera\_sim} package and applied it to the RFI simulation.
The hera\_sim package has already made a significant contribution to the simulation of RFI~\citep{kerrigan2019optimizing}. Both HERA and SKA1-LOW regard the detection of EOR as their important scientific goal and focus on the research of the frequency scope from 100 - 200 MHz where EOR is likely to be detected~\citep{de2018antenna,thyagarajan2017status}. 

We considered narrowband RFI and broadband RFI in the RFI simulation. The narrowband RFI is the primary RFI commonly caused by DTV, satellite communications, commercial FM. The broadband RFI is transient and strong, usually from the lightning and transmission cables. We added the two types of RFI with different ratios to the visibility data obtained in Section \ref{norfi} and generated three RFI-polluted measurement set (MS) files named Sim\_RFI-1, Sim\_RFI-2, and Sim\_RFI-3, respectively. The visibility functions of the three new MS files are stored in three Numpy arrays with the same shape as the original MS files. The details of these RFI-polluted files are given in Table \ref{tab:sim}. 

\begin{table*}
\centering

\caption{The information of the RFI-polluted data. Amp represents the amplitude of the RFI, the $\mu$ and $\sigma$ represent the mean and standard deviation of the overall data.}
\begin{tabular}{ccccccccc}
\toprule
Data set &  Sim\_RFI-1 & Sim\_RFI-2 & Sim\_RFI-3\\
\midrule
Phase-centre &RA=+15$^\circ$, Dec=-45$^\circ$ & RA=+30$^\circ$, Dec=-45$^\circ$ & RA=+45$^\circ$, Dec=-45$^\circ$\\
Observation time(2020-01-01) &  10:27 -  10:37 &  11:27 -  11:37 & 12:23 - 12:37   \\
RFI ratio & 10\% & 7.5\% & 1.1\% \\
RFI composition & 70\%Narrow band RFI + 30\%Broad band RFI  & 90\%Narrow band RFI+ 10\%Broad band RFI & 100\%Narrow band RFI \\
RFI strength &  $\mu$<Amp<$\mu$ + 13$\sigma$& $\mu$<Amp<$\mu$ + 15$\sigma$ & $\mu$<Amp<$\mu$ + 20$\sigma$\\

\bottomrule
\label{tab:sim}
\end{tabular}
\end{table*}

\subsection{Dataset}

Each MS file generated in Section \ref{norfi} and Section \ref{rfi} is 1.25GB in size. Hence the total size of the six MS files reaches 7.5GB. We selected the visibility data from the baseline SKA1-LOW020 to SKA1-LOW004. Each MS file has 300000 data pairs (amplitude, phase). Hence, the total number of data pairs we used are 300000 $\times$ 6.

We randomly selected 80$\%$  data pairs from the Sim\_RFI-2 as the training set and the left 20$\%$  as the testing set. The visibility data from the other two RFI-polluted MS files (i.e., the Sim\_RFI-1 and the Sim\_RFI-3) and the three original MS files are used to test the robustness of the model. Since the SKA1-LOW data are all simulated and generated, labeling the data is trivial.

\subsection{CNN Modelling}
\begin{figure}
    \centering
    \includegraphics[width=7 cm]{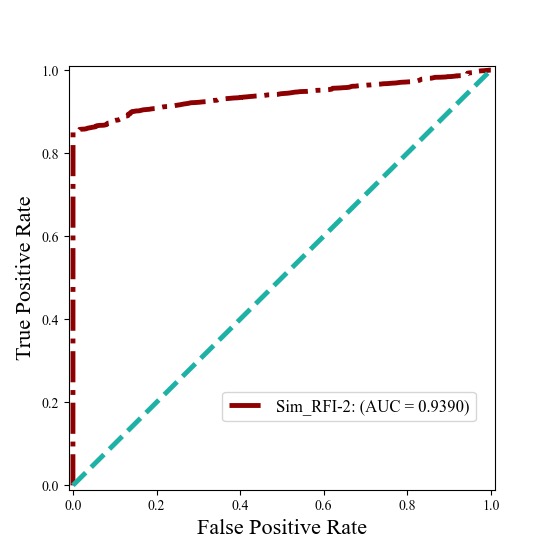}
    \caption{The ROC curve for training Sim\_RFI-2 data. The value of AUC is 0.9390.}
    \label{fig:ROC}
\end{figure}

Through training with the data of Sim\_RFI-2, we obtained a CNN model with the recall of 97.2\%, the precision of 99.1\%, and the accuracy of 99.8\%. The ROC curves are shown in Fig.~\ref{fig:ROC} .

\subsection{Results}

\begin{figure*}
\centering
\begin{minipage}[t]{0.48\textwidth}
\centering
\includegraphics[width=6cm]{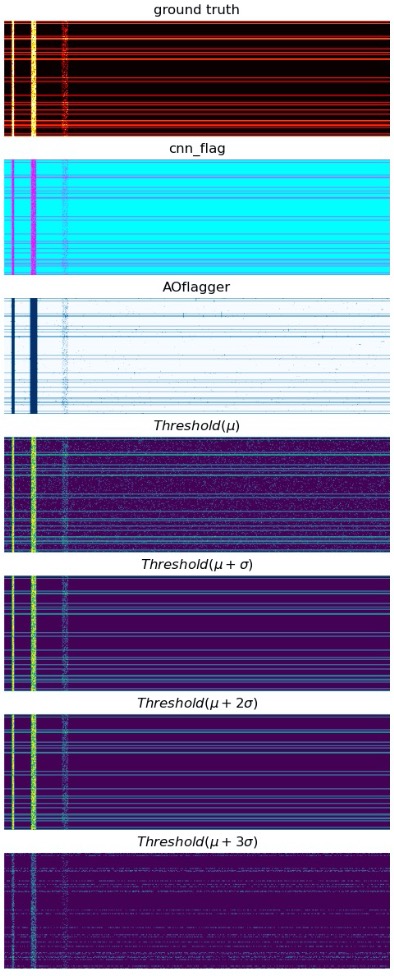}
\end{minipage}
\begin{minipage}[t]{0.50\textwidth}
\centering
\includegraphics[width=6cm]{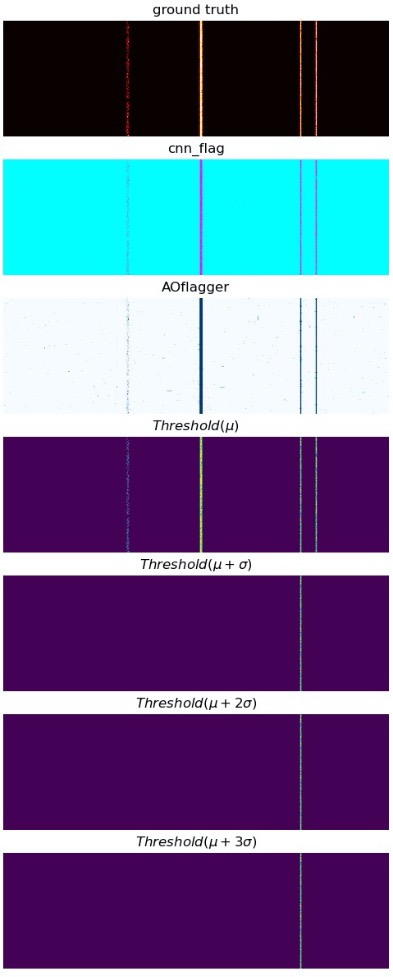}
\end{minipage}
\caption{Flagging results of Sim\_RFI-1~(left) and Sim\_RFI-3~(right) with different methods. In each panel, the horizontal axis is frequency, and the vertical axis is time. }
\label{fig:sim1_sim2}
\end{figure*}

We fed the two simulated RFI-polluted validation sets (Sim\_RFI-1 and Sim\_RFI-3) into the trained CNN model. The flagging results of our CNN model and AOflagger are presented in Fig.~\ref{fig:sim1_sim2}. The performance evaluation is given in Table \ref{tab:result}. Since the thresholding based on the three-sigma rule \citep{pukelsheim1994three} is often used in traditional data processing to identify outliers of data, we also added the flagging results under different thresholds for comparison. 

We also applied our CNN model and AOflagger to identify RFI in the dataset without RFI, i.e., the datasets of Sim-1, Sim-2, and Sim-3. The result shows that our CNN model detected no false-positive RFI. The average false-positive RFI ratio of AOflagger was 0.13\%. \citep{offringa2013lofar} estimated that misdetection of AOFlagger accounts for approximately 0.4\% of detections. Therefore, we think 0.13\% should be the misidentification of AOflagger.

\begin{table*}
\caption{Performance evaluation of CNN and AOflagger on the RFI-polluted validation data sets.}
\begin{tabular}{ccccccc}
\toprule

\multirow{3}{*}{Data Set} & \multicolumn{3}{c}{CNN} &  \multicolumn{3}{c}{AOFlagger}\\

\cline{2-7}
& Recall  & Precision  & Accuracy & Recall  & Precision  & Accuracy \\

\midrule
Sim\_RFI-1 &97.4\%&97.5\%&97.8\%  &100\%&88.2\%&98.7\%\\


Sim\_RFI-3    & 99.7\% & 99.4\%  & 99.8\%  & 99.8\% & 71.9\%  & 97.1\% \\ 

\bottomrule
\end{tabular}
\label{tab:result}
\end{table*}

\section{RFI Mitigation For Data from LOFAR and Meerkat}
\label{4}
We further applied our model to real observational data from LOFAR and MeerKAT. 

\subsection{RFI mitigation for data from LOFAR}
\label{lofar_rfi}

\begin{figure}
\begin{center}
	\includegraphics[width=5cm]{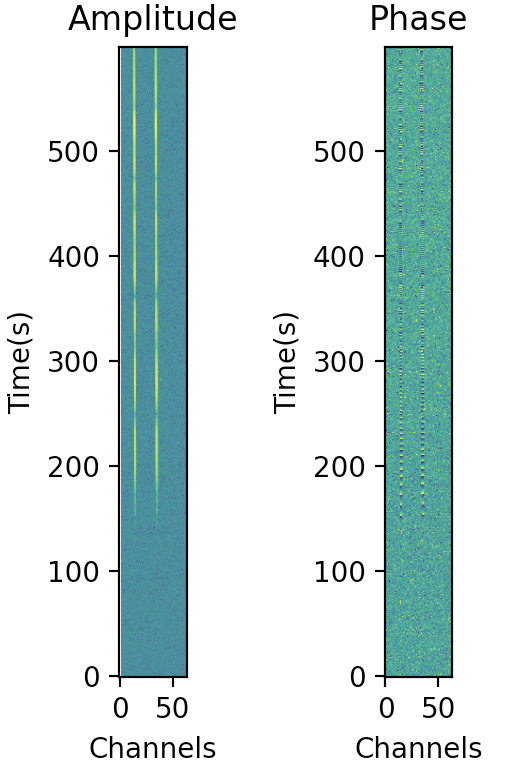}
\end{center}
    \caption{The visibility data of LOFAR-Set. The left row shows the amplitude value and the right row shows the phase value. The data pairs of the first 200 seconds are used for testing, and the left 400 seconds used for training.}
    \label{fig:lofar}
\end{figure}

LOFAR is a new-generation radio interferometer constructed in the north of the Netherlands and across Europe. Utilizing a novel phased array design, LOFAR covers the frequency range of  10 to 240 MHz and provides unique observing capabilities. We used its 10 minutes of visibility data, named as LOFAR-Set, for the following research. LOFAR-Set consists of 64 channels covering from 137.21MHz to 137.39MHz. Fig.~\ref{fig:lofar} shows the information of the amplitude and phase, in which RFI is easy to be found. We used the first 200 seconds, i.e., 12800 data pairs (amplitude, phase), for testing, and the other 400 seconds, i.e., 25600 data pairs (amplitude, phase), for training.

Although the RFI in Fig.~\ref{fig:lofar} is obvious, it is not marked or flagged. In others words, the dataset of LOFAR-Set is not labeled for CNN. Therefore, we used AOflagger with its default SumThreshold to flag the training dataset of the LOFAR-Set and took the flagging results as the reference truth. From the experimental results in Sec.3, AOFlagger cannot fully identify the RFI. Therefore, the labels flagged by AOflagger are not the ground truth but are still an essential reference for us.

\begin{figure*}
\begin{center}

	\includegraphics[width=16cm]{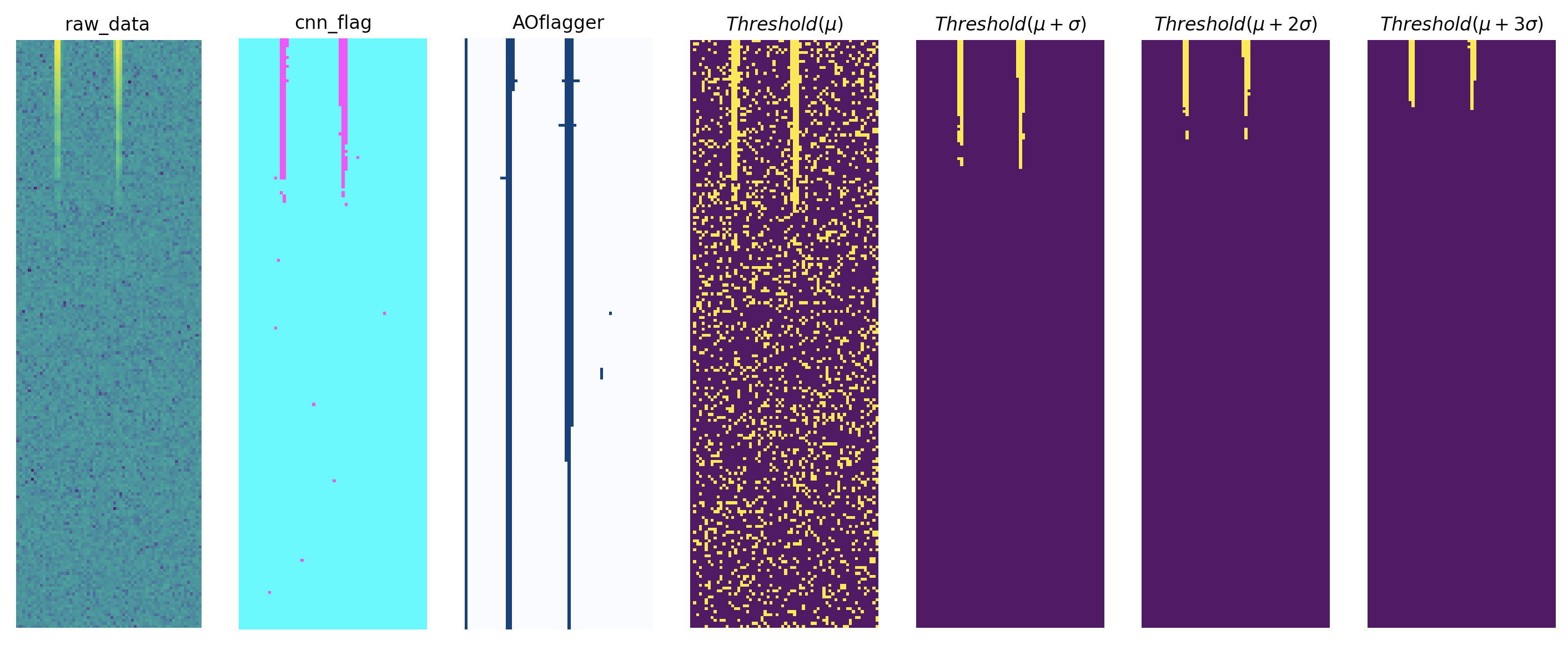}
\end{center}
    \caption{The results of the different RFI mitigation methods for LOFAR-Set. The horizontal axis is frequency, and the vertical axis is time. The first row is the amplitude of the testing data. The second and the third row are flagging results of our CNN method and the AOflagger, respectively. The others correspond to applying different thresholds to the data for flagging.}
    \label{fig:lofar_result}
\end{figure*}

The flagging results are shown in Fig.~\ref{fig:lofar_result}.

\subsection{RFI mitigation for data from Meerkat}

\begin{figure}
\begin{center}
	\includegraphics[width=8cm]{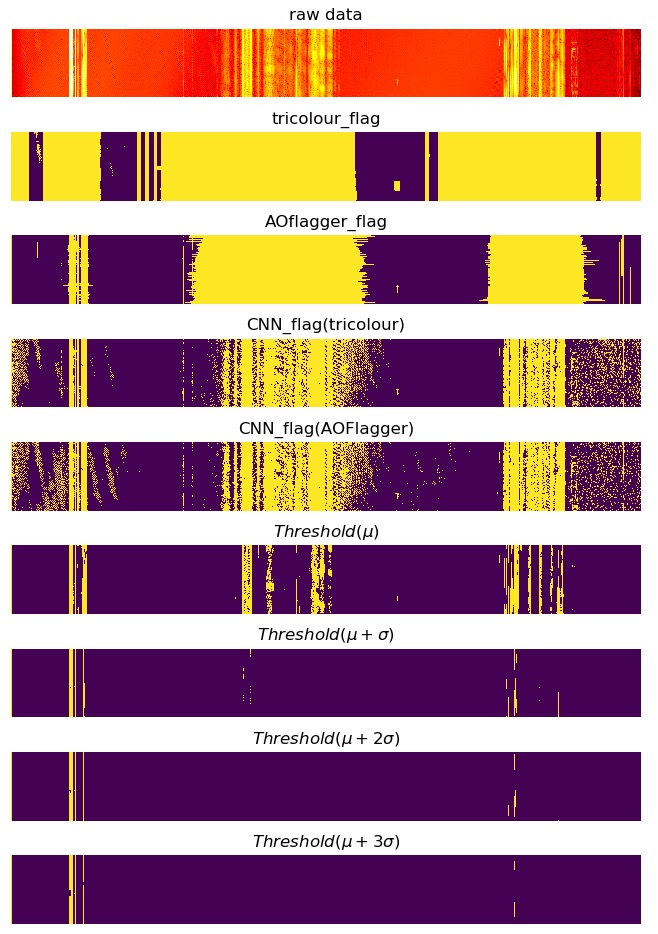}
\end{center}
    \caption{The results of the different RFI mitigation methods for MeerKAT-Set. The horizontal axis represents channels, and the vertical axis represents time. The first row shows the testing data, i.e., the target scan. The second and third rows are the flagging results of Tricolour and AOflagger, respectively. The fourth and fifth rows show the flagging results of our CNN model with the ground truth obtained from Tricolour and AOflagger, respectively. }
    \label{fig:meerkat}
\end{figure}

The MeerKAT radio telescope is a precursor of the SKA mid-frequency telescope, located in the arid Karoo region of the Northern Cape Province in South Africa. It comprises 64 antennas with a diameter of 13.5 meters. The observational data from MeerKAT we used, named MeerKAT-Set here, consists of a calibrator scan and a target scan, with a frequency band from 857.15MHz to 1.71GHz. The calibrator was scanned for 10 minutes, 4096 channels, so the number of data pairs is 303030. The target was scanned for 15 minutes, 1024 channels, so the number of data pairs is 114688. The 303030 data pairs are used for training, and the 114688 data pairs are used for testing. The top row in Fig.~\ref{fig:meerkat} is the map of the amplitude values of the target observation.

Like the dataset of LOFAR-Set, the dataset of MeerKAT-Set is not labeled either. We used AOflagger with default SumThreshold to flag the training dataset of MeerKAT-Set and set the results as the ground truth. In addition, MeerKAT commonly uses "Tricolour" for flagging\footnote{https://github.com/ska-sa/tricolour}. Tricolour is a specified AOFlagger version that can be run under the Dask framework~(\cite{Dask}). Tricolour uses a severe strategy for calibrator scan. We also used Tricolour to flag the training dataset of MeerKAT-Set and set the result as the ground truth.

The flagging results are shown in Fig.~\ref{fig:meerkat}.



\section{Discussion}
\label{5}
\subsection{Generalisation}

Due to the relatively small number of data samples, the ability to generalise the model cannot be verified very accurately yet. The model obtained by training the simulated dataset Sim\_RFI-2 shows satisfactory RFI identification on Sim\_RFI-1 and Sim\_RFI-3. In addition, the model adapts well to real observed datasets from LOFAR and MeerKat.
The current proposed CNN model has a certain generalization ability from these results, but more data are needed to test it.

\subsection{Comparison with AOFlagger}
AOFlagger uses a morphology-based RFI detection algorithm. It is more sensitive to changes in the data between adjacent channels, which easily allows AOflagger to identify as much RFI-contaminated data as possible.

Table \ref{tab:result} shows that for the simulated datasets, the accuracy of our CNN model and AOFlagger is almost the same. However, the precision of our model is higher than AOFlagger. By contrast, the recall of AOflagger is slightly higher. The possible reason is that we fed the training data into the CNN model for training in an unordered manner, which aligns with the feature that the time RFI happens has nothing to do with frequency. It is equivalent to reducing the redundancy of irrelevant features. 

\begin{figure*}
\begin{center}
	\includegraphics[width=16cm]{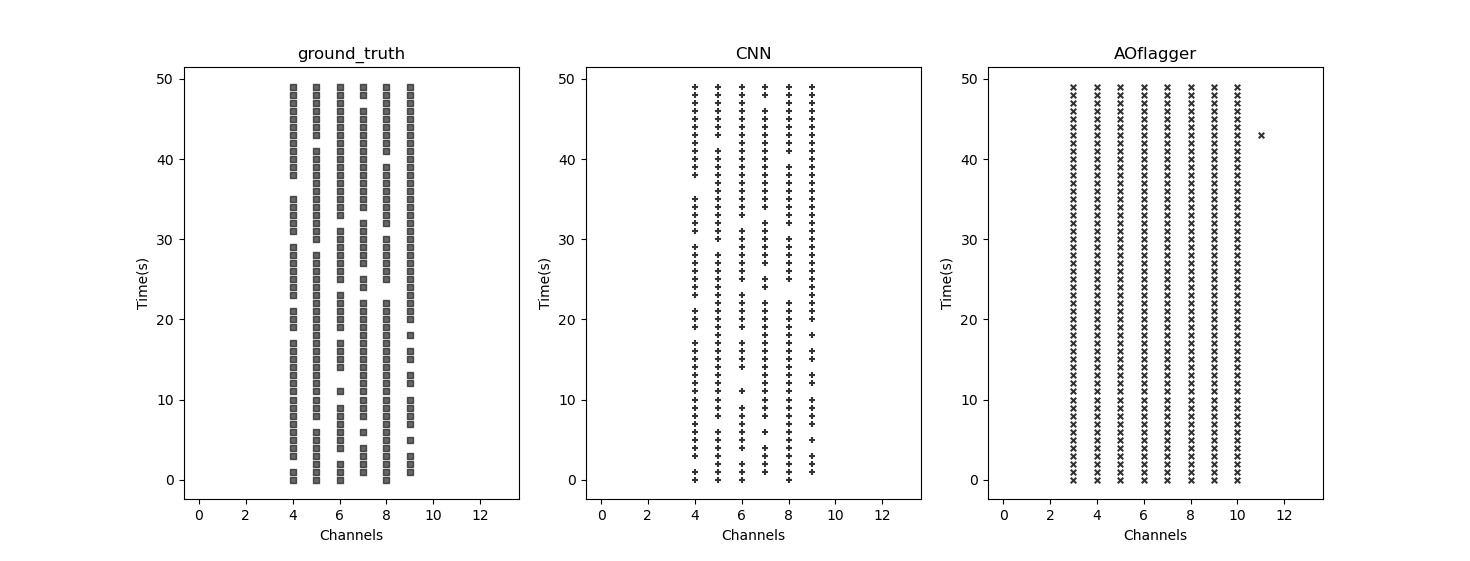}
\end{center}
    \caption{Zoomed-in diagram of Sim\_RFI-3 data set. AOFlagger incorrectly identifies the visibilities of both left and right edges.}
    \label{fig:edge}
\end{figure*}

Fig.~\ref{fig:edge} is a partial zoomed-in diagram of the Sim\_RFI-3 data set. It is evident that AOFlagger uses morphology to identify, as illustrated by the misidentification of the edges as RFI. Our proposed CNN-based recognition method can correctly identify the visibilities at the edges and can obtain a more reliable identification than AOFlagger.

\subsection{Weak RFI detection}

For some traditional RFI detection algorithms based on the changes of the adjacent signal, the intensity is one of the essential criteria for detection. Therefore, we also consider RFI with different intensities in our experiments.

We make a comparison of the traditional threshold method and our CNN model. The result (See Fig.~\ref{fig:sim1_sim2}) shows that there exists some difference between the two methods. Since the intensity of RFI is uncertain, RFI cannot be well identified by setting a definite threshold value. Our CNN model performs better on flagging different intensities of RFI on the simulation data (which has ground truth) than the threshold method, which indicates a potential advantage for weak RFI detection.

\subsection{RFI Identification Speed}
The speed of the proposed CNN-based RFI identification is slower than Aoflagger. We compared the running time between our CNN model and AOflagger. The result in Table \ref{tab:time} shows that AOflagger is nearly twice as fast as our CNN model except for the LOFAR-Set. The speed of AOflagger for LOFAR-Set is four times as fast as our model. Therefore, AOFlagger is suitable for real-time processing, and the detection method of CNN models is suitable for subsequent scientific studies.

\begin{table}
	\centering
	\caption{The running time comparison of CNN and AOflagger. Both of them use an Intel(R) Xeon(R) CPU E5-2620 v4 @ 2.10GHz.}
	\label{tab:time}
	\begin{tabular}{lccr} 
		\toprule
		Data set &  Time(CNN) & Time(AOflagger)\\
		\midrule
	    Sim\_RFI-1 & 31s & 18s \\
	    Sim\_RFI-2 & 31s & 17s\\
        Sim\_RFI-3 & 32s & 18s\\
		LOFAR-Set & 4s & 1s\\
		MeerKAT-Set & 142s & 98s\\
		\bottomrule
	\end{tabular}
	\label{tab:time}
\end{table}

\subsection{Limitations}

The CNN model obtained by training a part of the LOFAR-Set (200-600s) has partially different labeling results from AOflagger when identifying RFI in other parts of the LOFAR-Set. The possible reason is that we only used the default RFI tagging strategy of AOflagger.

The computational cost of our CNN model is unsatisfactory as per Table \ref{tab:time}. The possible reason is that our CNN model is built on Python-based PyTorch framework\citep{paszke2019pytorch}. Le-Net does not have a speed advantage over AOflagger, for the latter is based on C++ language. Much work for optimizing our implementation remains to be done.

\section{CONCLUSION}
\label{6}
RFI identification is an indispensable work for the data processing of next generation radio telescopes. This paper describes a method for simulating observations of SKA1-LOW by RASCIL and hera\_sim package to obtain reliable simulation data, designs a CNN model and applies it to the simulated data and the real data from Meerkat and LOFAR. The results show that our CNN model has a good performance in RFI identification. 

However, we realize that there are still some limitations with our CNN model. First, we need real ground truth to guide the training data to a well-performing model. Currently, we have to manually review the correctness of the flags used for the training data. Second, the Python-based Le-Net has limitations in its computational speed, which makes its efficiency unsatisfactory.

The random nature of the human activity and the geographical variability of telescopes make identifying RFI a difficult task. At the same time, weak signals such as the cosmic microwave background and EoR also place high demands on the accuracy of RFI identification. In future work, we will continue to refine the method for SKA1-LOW observation simulations and try to use GPU for our model or use parallel computing to boost the computing speed of our CNN model. We will also try to use reference antennas to get a more accurate ground truth to enhance the ability of our model to identify RFI.

\section*{Acknowledgements}
This work is supported by the National SKA Program of China (2020SKA0110300), the Joint Research Fund in Astronomy (U1931141,U1831204) under cooperative agreement between the National Natural Science Foundation of China (NSFC) and the Chinese Academy of Sciences (CAS), the Funds for International Cooperation and Exchange of the National Natural Science Foundation of China (11961141001), the National Natural Science Foundation of China (No.11903009),
the Innovation Research for the Postgraduates of Guangzhou University (2020GDJC-D20), Fundamental and Application Research Project of Guangzhou (202102020677)

This work is also supported by Astronomical Big Data Joint Research Center, co-founded by National Astronomical Observatories, Chinese Academy of Sciences and Alibaba Cloud. In addition, many thanks to all the members of the RASCIL development team.

 We do appreciate the anonymous referee for valuable and helpful comments and suggestions.
 
\section{Data Availability}

All the program code and part of the test data are stored in Github repository, the link is \url{https://github.com/astronomical-data-processing/CNN_RFI.git}.

\newpage



\bibliographystyle{mnras}
\bibliography{rfi} 

\begin{thebibliography}{}
\makeatletter
\relax
\def\mn@urlcharsother{\let\do\@makeother \do\$\do\&\do\#\do\^\do\_\do\%\do\~}
\def\mn@doi{\begingroup\mn@urlcharsother \@ifnextchar [ {\mn@doi@}
  {\mn@doi@[]}}
\def\mn@doi@[#1]#2{\def\@tempa{#1}\ifx\@tempa\@empty \href
  {http://dx.doi.org/#2} {doi:#2}\else \href {http://dx.doi.org/#2} {#1}\fi
  \endgroup}
\def\mn@eprint#1#2{\mn@eprint@#1:#2::\@nil}
\def\mn@eprint@arXiv#1{\href {http://arxiv.org/abs/#1} {{\tt arXiv:#1}}}
\def\mn@eprint@dblp#1{\href {http://dblp.uni-trier.de/rec/bibtex/#1.xml}
  {dblp:#1}}
\def\mn@eprint@#1:#2:#3:#4\@nil{\def\@tempa {#1}\def\@tempb {#2}\def\@tempc
  {#3}\ifx \@tempc \@empty \let \@tempc \@tempb \let \@tempb \@tempa \fi \ifx
  \@tempb \@empty \def\@tempb {arXiv}\fi \@ifundefined
  {mn@eprint@\@tempb}{\@tempb:\@tempc}{\expandafter \expandafter \csname
  mn@eprint@\@tempb\endcsname \expandafter{\@tempc}}}

\bibitem[\protect\citeauthoryear{Akeret, Chang, Lucchi  \& Refregier}{Akeret
  et~al.}{2017}]{akeret2017radio}
Akeret J.,  Chang C.,  Lucchi A.,   Refregier A.,  2017, Astronomy and
  computing, 18, 35

\bibitem[\protect\citeauthoryear{Baan}{Baan}{2011}]{baan2011rfi}
Baan W.~A.,  2011, in 2011 XXXth URSI General Assembly and Scientific
  Symposium. pp~1--2

\bibitem[\protect\citeauthoryear{Baan, Fridman  \& Millenaar}{Baan
  et~al.}{2004}]{baan2004radio}
Baan W.~A.,  Fridman P.,   Millenaar R.,  2004, The Astronomical Journal, 128,
  933

\bibitem[\protect\citeauthoryear{Bhowan, Johnston, Zhang  \& Yao}{Bhowan
  et~al.}{2012}]{bhowan2012evolving}
Bhowan U.,  Johnston M.,  Zhang M.,   Yao X.,  2012, IEEE Transactions on
  Evolutionary Computation, 17, 368

\bibitem[\protect\citeauthoryear{Booth \& Jonas}{Booth \&
  Jonas}{2012}]{booth2012overview}
Booth R.,  Jonas J.,  2012, African Skies, 16, 101

\bibitem[\protect\citeauthoryear{Bradley}{Bradley}{1997}]{bradley1997use}
Bradley A.~P.,  1997, Pattern recognition, 30, 1145

\bibitem[\protect\citeauthoryear{{Braun}}{{Braun}}{1996}]{Braun+1996}
{Braun} R.,  1996, in {Raimond} E.,  {Genee} R.,  eds,  Astrophysics and Space
  Science Library Vol. 208, The Westerbork Observatory, Continuing Adventure
  in. p.~167 (\mn@eprint {arXiv} {astro-ph/9512060}),
  \mn@doi{10.1007/978-94-009-1734-7\_9}

\bibitem[\protect\citeauthoryear{Cornwell, Braun  \& Briggs}{Cornwell
  et~al.}{1999}]{cornwell1999deconvolution}
Cornwell T.,  Braun R.,   Briggs D.~S.,  1999, in Synthesis Imaging in Radio
  Astronomy II. p.~151

\bibitem[\protect\citeauthoryear{{Dask Development Team}}{{Dask Development
  Team}}{2016}]{Dask}
{Dask Development Team} 2016, Dask: Library for dynamic task scheduling.
\url {https://dask.org}

\bibitem[\protect\citeauthoryear{Dewdney et~al.}{Dewdney
  et~al.}{2019}]{dewdney2019ska1}
Dewdney P.,  et~al., 2019, Technical report, SKA1 Design Baseline Description.
SKA-TEL-SKO-0001075, internal SKA document

\bibitem[\protect\citeauthoryear{El-Sawy, Hazem  \& Loey}{El-Sawy
  et~al.}{2016}]{el2016cnn}
El-Sawy A.,  Hazem E.-B.,   Loey M.,  2016, in International conference on
  advanced intelligent systems and informatics. pp 566--575

\bibitem[\protect\citeauthoryear{Hashemi}{Hashemi}{2012}]{6179821}
Hashemi H.,  2012, \mn@doi [IEEE Signal Processing Magazine]
  {10.1109/MSP.2012.2185897}, 29, 82

\bibitem[\protect\citeauthoryear{Ioffe \& Szegedy}{Ioffe \&
  Szegedy}{2015}]{DBLP:journals/corr/IoffeS15}
Ioffe S.,  Szegedy C.,  2015, CoRR, abs/1502.03167

\bibitem[\protect\citeauthoryear{Kerrigan et~al.,}{Kerrigan
  et~al.}{2019}]{kerrigan2019optimizing}
Kerrigan J.,  et~al., 2019, Monthly Notices of the Royal Astronomical Society,
  488, 2605

\bibitem[\protect\citeauthoryear{LeCun, Bottou, Bengio  \& Haffner}{LeCun
  et~al.}{1998}]{lecun1998gradient}
LeCun Y.,  Bottou L.,  Bengio Y.,   Haffner P.,  1998, Proceedings of the IEEE,
  86, 2278

\bibitem[\protect\citeauthoryear{Lonsdale et~al.,}{Lonsdale
  et~al.}{2009}]{lonsdale2009murchison}
Lonsdale C.~J.,  et~al., 2009, Proceedings of the IEEE, 97, 1497

\bibitem[\protect\citeauthoryear{Macquart et~al.,}{Macquart
  et~al.}{2010}]{macquart2010commensal}
Macquart J.-P.,  et~al., 2010, Publications of the Astronomical Society of
  Australia, 27, 272

\bibitem[\protect\citeauthoryear{Mellema et~al.,}{Mellema
  et~al.}{2013}]{mellema2013reionization}
Mellema G.,  et~al., 2013, Experimental Astronomy, 36, 235

\bibitem[\protect\citeauthoryear{Montavon, Orr  \& Müller}{Montavon
  et~al.}{2012}]{2012Efficient}
Montavon G.,  Orr G.~B.,   Müller K.,  2012, 10.1007/978-3-642-35289-8, 9

\bibitem[\protect\citeauthoryear{Nair \& Hinton}{Nair \&
  Hinton}{2010}]{10.5555/3104322.3104425}
Nair V.,  Hinton G.~E.,  2010, in Proceedings of the 27th International
  Conference on International Conference on Machine Learning. ICML'10.
Omnipress, Madison, WI, USA, p. 807–814

\bibitem[\protect\citeauthoryear{Offringa, de Bruyn, Zaroubi  \&
  Biehl}{Offringa et~al.}{2010}]{offringa2010lofar}
Offringa A.,  de Bruyn A.,  Zaroubi S.,   Biehl M.,  2010, arXiv preprint
  arXiv:1007.2089

\bibitem[\protect\citeauthoryear{Offringa et~al.,}{Offringa
  et~al.}{2013}]{offringa2013lofar}
Offringa A.,  et~al., 2013, Astronomy \& astrophysics, 549, A11

\bibitem[\protect\citeauthoryear{Offringa et~al.,}{Offringa
  et~al.}{2015}]{offringa2015low}
Offringa A.,  et~al., 2015, Publications of the Astronomical Society of
  Australia, 32

\bibitem[\protect\citeauthoryear{Offringa, Mertens  \& Koopmans}{Offringa
  et~al.}{2019}]{offringa2019impact}
Offringa A.,  Mertens F.,   Koopmans L.,  2019, Monthly Notices of the Royal
  Astronomical Society, 484, 2866

\bibitem[\protect\citeauthoryear{Parsons et~al.,}{Parsons
  et~al.}{2014}]{parsons2014new}
Parsons A.~R.,  et~al., 2014, The Astrophysical Journal, 788, 106

\bibitem[\protect\citeauthoryear{Paszke et~al.,}{Paszke
  et~al.}{2019}]{paszke2019pytorch}
Paszke A.,  et~al., 2019, Advances in neural information processing systems,
  32, 8026

\bibitem[\protect\citeauthoryear{Perley \& Cornwell}{Perley \&
  Cornwell}{2003}]{perley2003evla}
Perley R.,  Cornwell T.,  2003

\bibitem[\protect\citeauthoryear{Pukelsheim}{Pukelsheim}{1994}]{pukelsheim1994three}
Pukelsheim F.,  1994, The American Statistician, 48, 88

\bibitem[\protect\citeauthoryear{Sutskever, Martens, Dahl  \& Hinton}{Sutskever
  et~al.}{2013}]{sutskever2013importance}
Sutskever I.,  Martens J.,  Dahl G.,   Hinton G.,  2013, in International
  conference on machine learning. pp 1139--1147

\bibitem[\protect\citeauthoryear{Thyagarajan}{Thyagarajan}{2017}]{thyagarajan2017status}
Thyagarajan N.,  2017, Proceedings of the International Astronomical Union, 12,
  102

\bibitem[\protect\citeauthoryear{Vafaei~Sadr, Bassett, Oozeer, Fantaye  \&
  Finlay}{Vafaei~Sadr et~al.}{2020}]{vafaei2020deep}
Vafaei~Sadr A.,  Bassett B.~A.,  Oozeer N.,  Fantaye Y.,   Finlay C.,  2020,
  Monthly Notices of the Royal Astronomical Society, 499, 379

\bibitem[\protect\citeauthoryear{Wayth et~al.,}{Wayth
  et~al.}{2015}]{wayth2015gleam}
Wayth R.,  et~al., 2015, Publications of the Astronomical Society of Australia,
  32

\bibitem[\protect\citeauthoryear{Wilson \& Martinez}{Wilson \&
  Martinez}{2003}]{wilson2003general}
Wilson D.~R.,  Martinez T.~R.,  2003, Neural networks, 16, 1429

\bibitem[\protect\citeauthoryear{Xie, Wang, Wei, Wang  \& Tian}{Xie
  et~al.}{2016}]{xie2016disturblabel}
Xie L.,  Wang J.,  Wei Z.,  Wang M.,   Tian Q.,  2016, in Proceedings of the
  IEEE Conference on Computer Vision and Pattern Recognition. pp 4753--4762

\bibitem[\protect\citeauthoryear{Yatawatta}{Yatawatta}{2021}]{yatawatta2021polarization}
Yatawatta S.,  2021, in 2020 28th European Signal Processing Conference
  (EUSIPCO). pp 1961--1965

\bibitem[\protect\citeauthoryear{Yuan, Bai, Jiao  \& Liu}{Yuan
  et~al.}{2012}]{yuan2012offline}
Yuan A.,  Bai G.,  Jiao L.,   Liu Y.,  2012, in 2012 10th IAPR International
  Workshop on Document Analysis Systems. pp 125--129

\bibitem[\protect\citeauthoryear{de Lera~Acedo, Pienaar  \&
  Fagnoni}{de~Lera~Acedo et~al.}{2018}]{de2018antenna}
de Lera~Acedo E.,  Pienaar H.,   Fagnoni N.,  2018, in 2018 International
  Conference on Electromagnetics in Advanced Applications (ICEAA). pp 636--639

\bibitem[\protect\citeauthoryear{de Vos, Gunst  \& Nijboer}{de~Vos
  et~al.}{2009}]{de2009lofar}
de Vos M.,  Gunst A.~W.,   Nijboer R.,  2009, Proceedings of the IEEE, 97, 1431

\makeatother
\end{thebibliography}









\bsp	
\label{lastpage}
\end{document}